\documentclass[a4paper]{jpconf}
\usepackage{graphicx}
\usepackage[usenames,dvipsnames]{color}
\usepackage{amsmath}
\usepackage{colordvi,epsf}
\usepackage{braket}
\usepackage{xcolor}
\begin{document}

\title{B-T phase diagram of Pd/Fe/Ir(111) computed with parallel tempering Monte Carlo}

\author{M B\"ottcher $^{1,2,3}$, S Heinze$^2$, S Egorov$^{4,5}$, J Sinova$^{1,6}$ and B Dup\'e$^{1,2}$}
\address{$^1$ Institut f\"ur Physik, Johannes Gutenberg Universit\"at Mainz, 55099 Mainz, Germany}
\address{$^2$ Institute of Theoretical Physics and Astrophysics, University of Kiel, 24098 Kiel, Germany}
\address{$^3$ Graduate School Materials Science in Mainz, 55128 Mainz, Germany}
\address{$^4$ Department of Chemistry, University of Virginia, Charlottesville, Virginia 22901, USA }
\address{$^5$ Leibniz-Institut f\"ur Polymerforschung, Institut Theorie der Polymere, 01069 Dresden, Germany}
\address{$^6$ Institute of Physics, Academy of Sciences of the Czech Republic, Cukrovarnick\'{a} 10, 162 53 Praha 6 Czech Republic}

\ead{m.boettcher@uni-mainz.de}

\begin{abstract}
{We use an atomistic spin model derived from density functional theory calculations for the ultra-thin film Pd/Fe/Ir(111) to show that temperature induces coexisting non-zero skyrmion and antiskyrmion densities.
We apply the parallel tempering Monte Carlo method in order to reliably compute thermodynamical quantities and the $B$-$T$ phase diagram in the presence of frustrated exchange interactions.
We evaluate the critical temperatures using the topological susceptibility.
We show that the critical temperatures depend on the magnetic field in contrast to previous work.
In total, we identify five phases: spin spiral, skyrmion lattice, ferromagnetic phase, intermediate region with finite topological charge and paramagnetic phase.
To explore the effect of frustrated exchange interactions, we calculate the $B$-$T$ phase diagram, when only effective exchange parameters are taken into account. 
}
\end{abstract}

\section{Introduction}

Chiral magnetic spin structures, such as magnetic skyrmions \cite{Bogdanov1994},
have received a lot of interest since
they are possible candidates as bits in data storage \cite{yu2012,Fert2013,Sampaio2013,Jiang2015a,Yu2017}.
Magnetic skyrmions are localized non-collinear spin-textures with a unique rotational sense which defines their chirality. 
The winding of the magnetization can be described by an integer topological charge which is called skyrmion number $Q$ \cite{Nagaosa2013}: 
\begin{eqnarray}
Q = \frac{1}{4\pi} \int \mathbf{m}\cdot\left(\frac{\partial \mathbf{m}}{\partial x}\times\frac{\partial \mathbf{m}}{\partial y}\right)\mathrm {d}x\mathrm {d}y\,,
\label{eq:sky_windingnumber}
\end{eqnarray}
where $x$ and $y$ are the spatial coordinates and $\mathbf{m}$ the unit vector of the magnetization.

Micromagnetic models have predicted the occurrence of magnetic skyrmions in  condensed matter \cite{Bogdanov1994,Bogdanov1989}.
Density functional theory (DFT) calculations have demonstrated that skyrmion stabilization can be explained by the competition between the Dzyaloshinskii-Moriya interaction (DMI) and the magnetic exchange interaction beyond the first nearest neighbor approximation \cite{Dupe2014,Simon2014,Dupe2016b}. The latter is enough to stabilize topologically protected states \cite{Leonov2015b,vonMalottki2017}. Then, the stability of skyrmions and antiskyrmions \cite{Dupe2016a,Hoffmann2017} is enhanced only by DMI \cite{Dzyaloshinsky1958,Moriya1960a}, which occurs when structural inversion symmetry is broken such as in chiral magnets or at surfaces or interfaces \cite{Fert1980,Crepieux1998a,Heinze2011}. 

In order to use magnetic skyrmions in data storage devices, the knowledge of their temperature ($T$) and magnetic field ($B$) dependence, which can be summarized in a $B$-$T$ phase diagram, is of great importance.
$B$-$T$ phase diagrams including skyrmion magnetic textures were first measured for MnSi bulk \cite{Muhlbauer2009a}, Fe$_{0.5}$Co$_{0.5}$Si thin-films \cite{Yu2010} and in the multiferroics Cu$_2$OSeO$_3$ \cite{Seki2012}.
In the case of MnSi, the $B$-$T$ phase diagram was reproduced via Monte Carlo (MC) simulations \cite{PhysRevB.88.195137}.
All these diagrams show long ranged ordered phases (spin spiral, ferromagnetic (FM), skyrmion lattice (SkX)), a short range ordered phase and a paramagnetic (PM) phase
which are separated by critical temperatures $T_\textup{c}$ which depend on the external magnetic field $B$. In none of the above mentioned cases, the topological charge has been measured. 
Recently, several works have been devoted to the analysis of the modified topological charge of skyrmions as a function of temperature and magnetic field. 
The topological properties of skyrmions have been analyzed at $T$ = 0 K in the context of electron transport \cite{Zhang2016j}.
A minimum model containing frustration of exchange interaction has been considered, showing a liquid phase of skyrmions and antiskyrmions \cite{Lin2016a}.
The temperature dependence of the topological charge of skyrmions was explored via MC simulations with and without impurities \cite{Silva:2014aa,Ambrose2013}, and shows decreasing charge at high temperature due to the melting of the skyrmion phase. 
A general $B$-$T$ phase diagram of a two dimensional film of a chiral magnet obtained by MC simulations showed the occurrence of topological charge at high magnetic field in the intermediate region far from the spin spiral ground state \cite{Hou2017a}. 

Skyrmions are also present in Fe (ultra-)thin films, e.g., Fe monolayer on Ir(111) \cite{Heinze2011}, Pd/Fe bilayer on Ir(111) \cite{Dupe2014,Simon2014,Romming2013,Romming2015,rozsa2016}, 3 monolayers of Fe on Ir(111) \cite{Hsu2016a} and Co ultrathin film Co/Ru(0001) \cite{Herve2018}.
In that case, magnetic interactions can be tuned by the choice of the magnetic film \cite{Belabbes2016a,Nandy2016}, the hybridization with the different substrates \cite{Dupe2016b,Belabbes2016a,Hardrat2009,PhysRevLett.99.187203}, or optional overlayers \cite{Dupe2014}.
Pd/Fe/Ir(111) shows isolated skyrmions which can be created and annihilated by the spin polarized current induced by the spin-polarized scanning tunneling microscope tip \cite{Romming2013}.
However, a $B$-$T$ phase diagram has not been obtained experimentally yet.

It was shown, that magnetic exchange interaction beyond the first nearest neighbor approximation without dipole-dipole interaction \cite{Dupe2014,Simon2014,vonMalottki2017} or first nearest exchange magnetic interaction with dipole-dipole interaction \cite{Leonov2016a} was necessary to model its magnetic states.
In particular, we have shown that the presence of frustrated exchange strongly enhance the barrier for skyrmion collapse into the FM state \cite{vonMalottki2017}.
A $B$-$T$ phase diagram of Pd/Fe/Ir(111) based on a Metropolis MC simulation has been reported by R\'ozsa \textit{et al.} \cite{rozsa2016}.
At low temperature, the system transits from a spin spiral phase to a SkX phase and finally to a FM phase with increasing magnetic field \cite{rozsa2016}, in agreement with previous works \cite{Dupe2014,Simon2014,Romming2013}.
Between the ordered phases and the PM phase, a fluctuation-disordered phase was found where the skyrmion lifetime is finite \cite{PhysRevB.88.195137,Bauer2013,Janoschek2013}.
In this study, the critical temperatures between the ordered phases and the fluctuation-disordered phase are independent of
the magnetic field \cite{rozsa2016} which differs from previous experimental work \cite{Muhlbauer2009a,Seki2012,PhysRevB.88.195137}.

Here, we study the temperature dependence of the contributions of skyrmion density (SkD, positive topological charge $Q_+$) and the antiskyrmion density (ASkD, negative topological charge $Q_-$) to the net topological charge $Q$. $Q_\pm$ are now real numbers which describe the two topological charge contributions in the whole super cell but cannot identify a particular skyrmion number \cite{PhysRevB.39.7212}. We use parallel tempering MC (PTMC) \cite{Swendsen1986,Hukushima:1996aa} to compute $Q_\pm$ to determine the $B$-$T$ phase diagram of Pd/Fe/Ir(111). We show that $Q_\pm$ increases with temperature and the corresponding topological susceptibility $\chi_{Q_\pm}$, which measures the correlations and fluctuations, can be used to obtain the critical temperatures in the whole range of magnetic fields.
Here, we particularly focus on the effect of the influence of frustration of magnetic exchange interaction on the critical temperature by computing the $B$-$T$ phase diagram with a full set of exchange coefficients obtained by DFT ($J_\mathrm{DFT}$) which contains magnetic exchange up to the 9th nearest neighbor and an effective exchange parameter ($J_\mathrm{eff}$) \cite{vonMalottki2017}.

\section{Computational details}

To obtain the $B$-$T$ phase diagram of Pd/Fe/Ir(111), we started from the electronic structure of the system to calculate the interactions between the magnetic moments with DFT calculations obtained in Refs.~\cite{Dupe2014,vonMalottki2017}. We then use these parameters in our MC simulation in order to calculate the temperature dependence of thermodynamic quantities. 

\subsection{Model}

We consider an ultra-thin film built from a monolayer of Pd in fcc-stacking on a monolayer of Fe in fcc stacking on an Ir(111) surface.
For this system, DFT calculations were carried out to determine the magnetic interactions.
The parameters were obtained with the FLEUR ab initio package \cite{PhysRevB.69.024415,Heide2009,fleur}.
Our spin models are based on the extended Heisenberg model:
\begin{eqnarray}
H = - \sum_{ij} J_{ij} (\mathbf m_i \cdot \mathbf m_j)
     -  \sum_{ij} \mathbf D_{ij} \cdot (\mathbf m_i \times \mathbf m_j) 
    + \sum_i K (m_i^z)^2
     -  \sum_i \mu_s \mathbf B \cdot \mathbf m_i\,,
\label{eq:modelH}
\end{eqnarray}
where $\mathbf m_i$ is the unit vector of the magnetization of the Fe atoms at sites $\mathbf{R}_i$, $J_{ij}$ are the magnetic exchange coefficients, $\mathbf{D}_{ij}$ is the DMI-vector, $K$ the coefficient of the magnetocrystalline anisotropy and $\mathbf{B}$ the magnetic field perpendicular to the film. The absolute value of the magnetization is $\mu_s$ = 3 $\mu_\mathrm{B}$, where $\mu_\mathrm{B}$ is the Bohr magneton.
{The} exchange constants $J_\mathrm{DFT} = \{J_n\}$ between the $n^{\mathrm{th}}$ nearest-neighbors in the {Fe} layer ${J}_1 \cdots {J}_{9}$ are given, in {meV}, by $14.40$, $-2.48$, $-2.69$, $0.52$, $0.74$, $0.28$, $0.16$, $-0.57$, $-0.21$, respectively.
{The} anisotropy constant is given by ${K}=-0.7$ me{V}, which corresponds to an out-of-plane easy axis and the {DMI} is restricted to nearest neighbors with a value given by ${D}=1.0$ me{V} and its direction is obtained in agreement with the model in {R}ef.~\cite{Fert1980}.
We use these DFT parameters unless stated otherwise.
When only an effective nearest neighbor exchange parameter is taken into account $J_\mathrm{eff}$ = 3.68 meV, ${K}=-0.7$ me{V} and ${D}=1.39$ me{V}.
Note, that these values are very close to those obtained based on experiments \cite{Romming2015,Hagemeister2015}.
For further details see {R}ef. \cite{Dupe2014,vonMalottki2017}.
The atomistic spin simulations were performed in a supercell with $N$ = (100$\times$100) magnetic moments, where we used periodic boundary conditions in the horizontal directions. This corresponds to a supercell of (27$\times$23.4 nm$^2$) where 6 spin spiral periods are stabilized.

\subsection{Methods}

To study the temperature dependence of Pd/Fe/Ir(111), we perform PTMC simulations to handle the occurrence of metastable states.
PTMC makes use of the temperature to exit metastable states that occur when frustration of exchange is present and isolated skyrmions or antiskyrmions appear \cite{Dupe2016a,PhysRevB.95.094423}.
We used 240 spin configurations (replicas), which are simultaneously simulated at different temperatures distributed in a geometric temperature set \cite{Katzgraber:2006aa}.
The simulations were initialized with the ground state ($T$ $\approx$ 0 K) configurations. The replicas of adjacent temperatures were swapped $10^4$ times (average steps). In between these swapping steps, the spin configurations were thermalized with $10^6$ Metropolis MC steps.
We calculate the total energy $E$ (according to equation~\ref{eq:modelH}) and the magnetization density $M$ ($M=|\sum \mathbf{m}_i|/N$).
In addition, we calculate the topological charge $Q$, which is defined as the difference between the absolute values of the SkD and ASkD contributions $Q_+$ and $Q_-$, which are obtained by taking only the positive or negative part of the integrand in Eq.~(\ref{eq:sky_windingnumber}) \cite{Berg1981}, respectively.

To distinguish between the different phases, we define the susceptibilities associated with the above mentioned quantities:

We define the heat capacity $C$ as
\begin{equation}
C = \frac{ \left( \braket{{E}^2}-\braket{E}^2 \right)}{k_\textup{B}T^2}
\end{equation}
and the magnetic susceptibility as:
\begin{equation}
\chi_M= \frac{\left(\braket{{M}^2}-\braket{M}^2\right)}{k_\textup{B}T}
\end{equation}
where $k_\textup{B}$ is the Boltzmann constant and $T$ is the temperature.

In analogy with the number of particles in the grand canonical ensemble \cite{Pathria2011}, we define the topological susceptibility as
\begin{equation}
\chi_{Q_\pm}=\frac{\left(\braket{{Q_\pm}^2}-{\braket{Q_\pm}}^2\right)}{{\braket{Q_\pm}}k_\textup{B}T }
\end{equation}
which describes the fluctuations and correlations of the SkD and ASkD.

Each order parameter is modeled by an arctangent function with a linear noise:
\begin{equation}
f(T) = \frac{I}{\pi} \arctan \left( \frac{2 \left(T-T_\textup{c} \right)}{w} \right)
\end{equation}
where $I$ is the intensity (area underneath the peak),
$T_\textup{c}$ the critical temperature and $w$ the mean height width.
This simple model allows a precise evaluation of the critical temperature by fitting the different susceptibilites with the derivative of the arctan function i.e. a Lorentzian function of the type:
\begin{equation}
f'(T) = \frac{I}{\pi} \frac{2w}{ 4 \left(T-T_\textup{c} \right)^2 +w^2}
\end{equation}

This fitting procedure leads to a precision of around 10\% on $T_{\mathrm{c}}$.

\section{Results}

\subsection{Stability diagram at low temperature}

\begin{figure}[thp]
\centering
\includegraphics[width=0.9\linewidth]{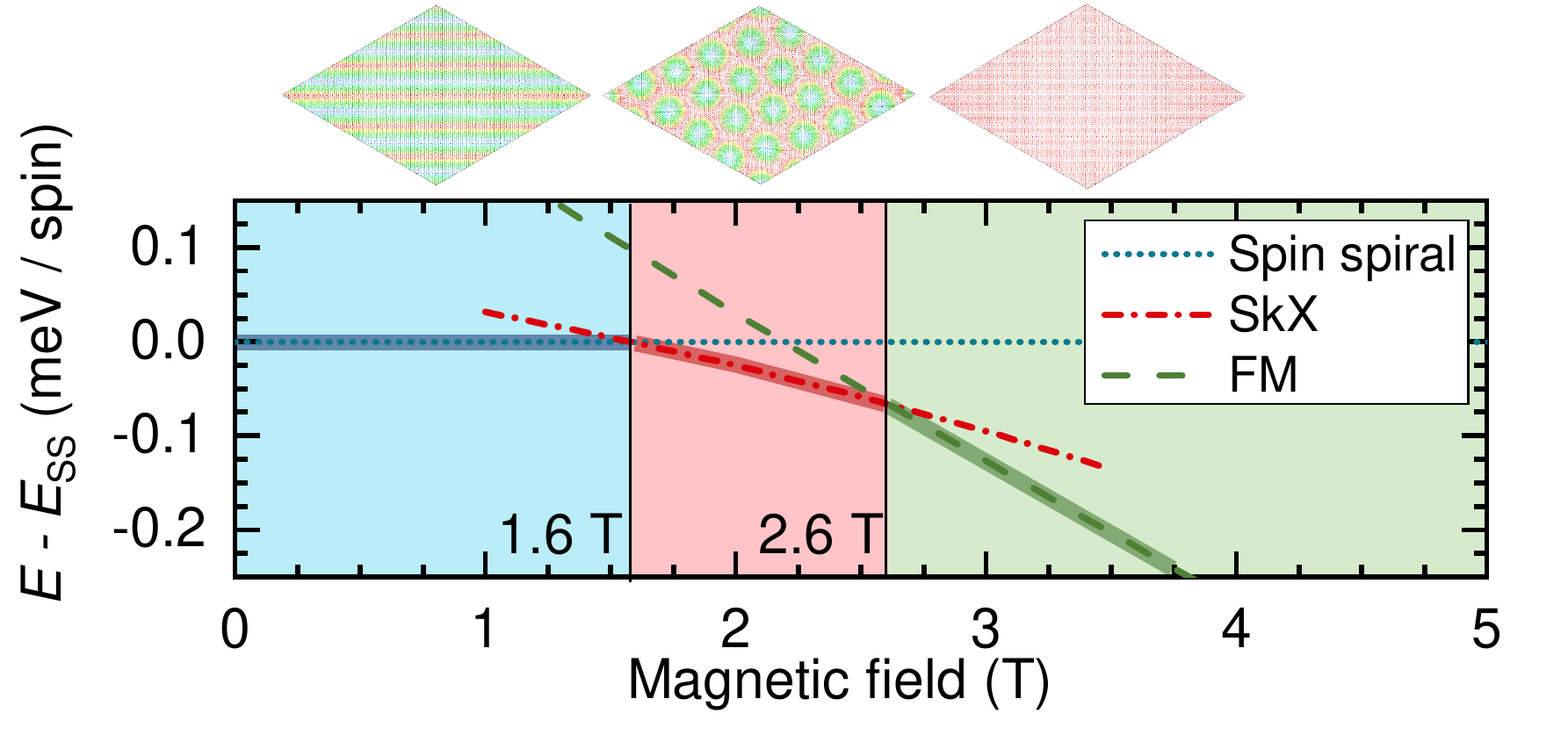}
\caption{{\bf Low temperature stability diagram of Pd(fcc)/Fe/Ir(111).}
Energy of the three different states as a function of the magnetic field at zero temperature: spin spiral (dotted blue line), SkX (chained red line), and FM state (dashed green line).
The colored areas mark the different ground state areas. $E_\mathrm{SS}$ is the energy of the spin spiral.
The plots ontop of the different ground state areas show the corresponding magnetic structures, with a color code from red (out of the plane) over green (in plane) to blue (into the plane).
}
\label{fig:stabilitydiagram_Pd-Fe-Ir111}
\end{figure}

First, we have performed standard Metropolis MC simulations of a spin spiral, a SkX containing 21 skyrmions and a FM state at $T$ = 0.001 K to obtain the stability diagram in figure~\ref{fig:stabilitydiagram_Pd-Fe-Ir111}. This stability diagram shows the most stable phase for different magnetic fields at 0 K. The magnetic structures can be seen in the plots above the corresponding ground state area in figure~\ref{fig:stabilitydiagram_Pd-Fe-Ir111}.
At low magnetic fields up to $B_\textup{s,1}$ = 1.6 T, the system has a spin spiral ground state in agreement with DFT.
The spin spiral propagates along the $\bar{\Gamma}-\bar{\textup{K}}$ direction with a wavelength of $\lambda$ = 4.5 nm which compares well with $\lambda$ $\approx$ 5-7 nm obtained experimentally \cite{Romming2013,Kubetzka2017}. 
Between 1.6 T and $B_\textup{s,2}$ = 2.6 T we find a stable SkX. The energy gain of the SkX state compared to the spin spiral or FM state is up to 0.04 meV per spin. The ordering of the skyrmions in the SkX along lines tilted to the lattice vectors was found with a first initial PTMC simulation at a magnetic field with SkX ground state. This previous PTMC simulation finds the most stable skyrmion lattice, which then can be used as the starting configuration for the PTMC simulation to obtain the thermodynamical quantities.
At magnetic fields above 2.6 T the FM phase is the ground state. These field values are in good agreement with previous MC simulations \cite{rozsa2016} and experimental measurements \cite{Romming2013}.

\subsection{Energy and heat capacity}

\begin{figure}[thp]
\centering
\includegraphics[width=0.9\linewidth]{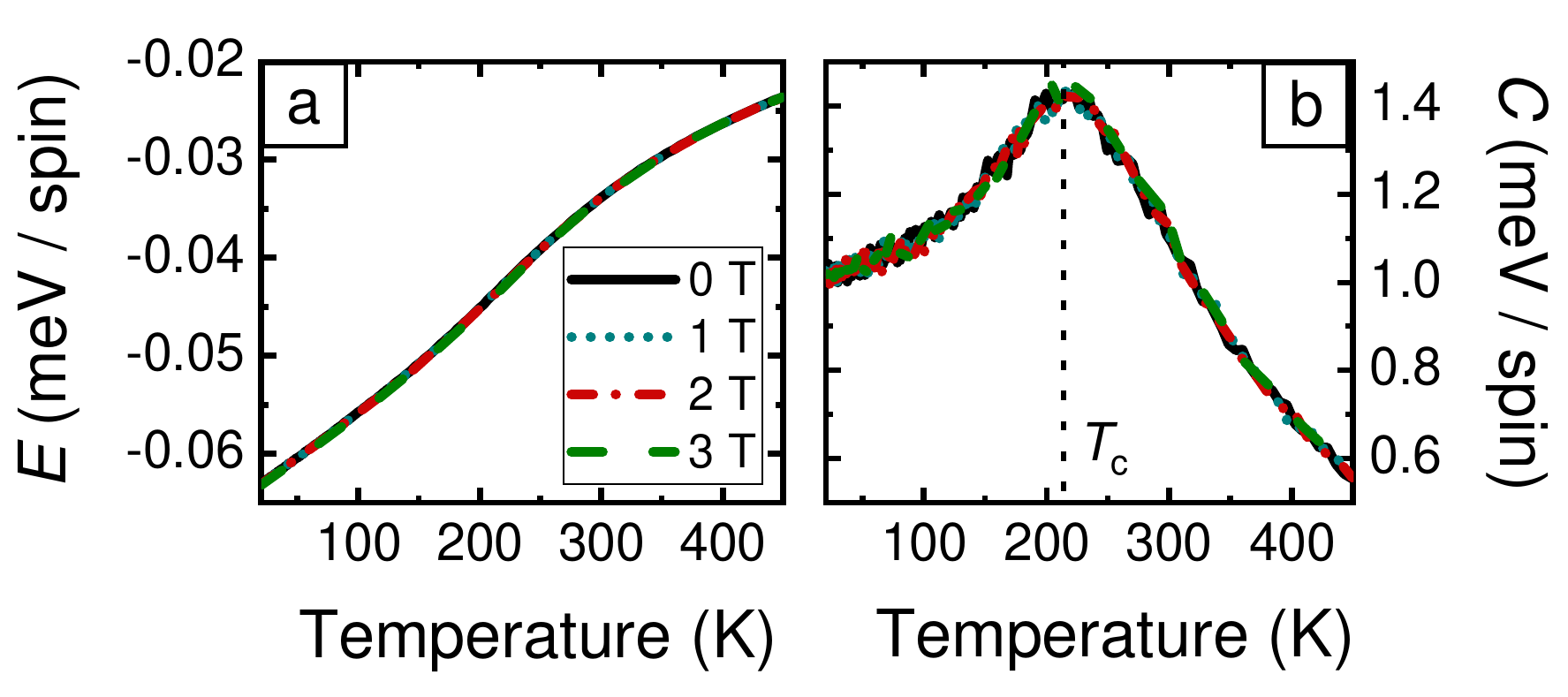}
\caption{{\bf Energy \textit{E} and heat capacity \textit{C} at selected magnetic fields.}
(a) Energy $E$ vs. temperature. The energy differences between the different fields can be seen in figure~\ref{fig:stabilitydiagram_Pd-Fe-Ir111}.
(b) Heat capacity vs. temperature. The dotted line defines the critical temperature $T_\textup{c}$ between the long or short range ordered phase to the disordered phase.}
\label{fig:Energy-heatcapacity}
\end{figure}

To identify the transition temperature from the ordered, long or short range, to the PM phase we use the peak in the heat capacity $C$ as in the work of Buhrandt \textit{et al.} \cite{PhysRevB.88.195137}.
Therefore, the temperature dependence of the total energy $E$ and the heat capacity $C$ for selected magnetic fields with different ground states (according to the stability diagram in figure~\ref{fig:stabilitydiagram_Pd-Fe-Ir111}) is shown in figure~\ref{fig:Energy-heatcapacity} (a) and (b), respectively.
For all magnetic fields the energy $E$ in figure~\ref{fig:Energy-heatcapacity} (a) increases with temperature as expected.
To obtain the inflection point of this increase, we calculate the heat capacity $C$, which shows a peak for all magnetic fields (see figure~\ref{fig:Energy-heatcapacity} (b)).
The field dependence of the energy and the heat capacity is minimal, as seen in figures~\ref{fig:stabilitydiagram_Pd-Fe-Ir111} (a) and (b). We do not observe the first order peak in $C$ that is characteristic for the Brazovskii scenario in MnSi as we have a two dimensional magnet with strong anisotropy and not a three dimensional magnet without anisotropy \cite{Brazovskii1975,PhysRevB.88.195137}.
We determine a critical temperature at about $T_\textup{c}$ $\approx$ 214 K.

\subsection{Magnetization}

\begin{figure}[thp]
\centering
\includegraphics[width=0.9\linewidth]{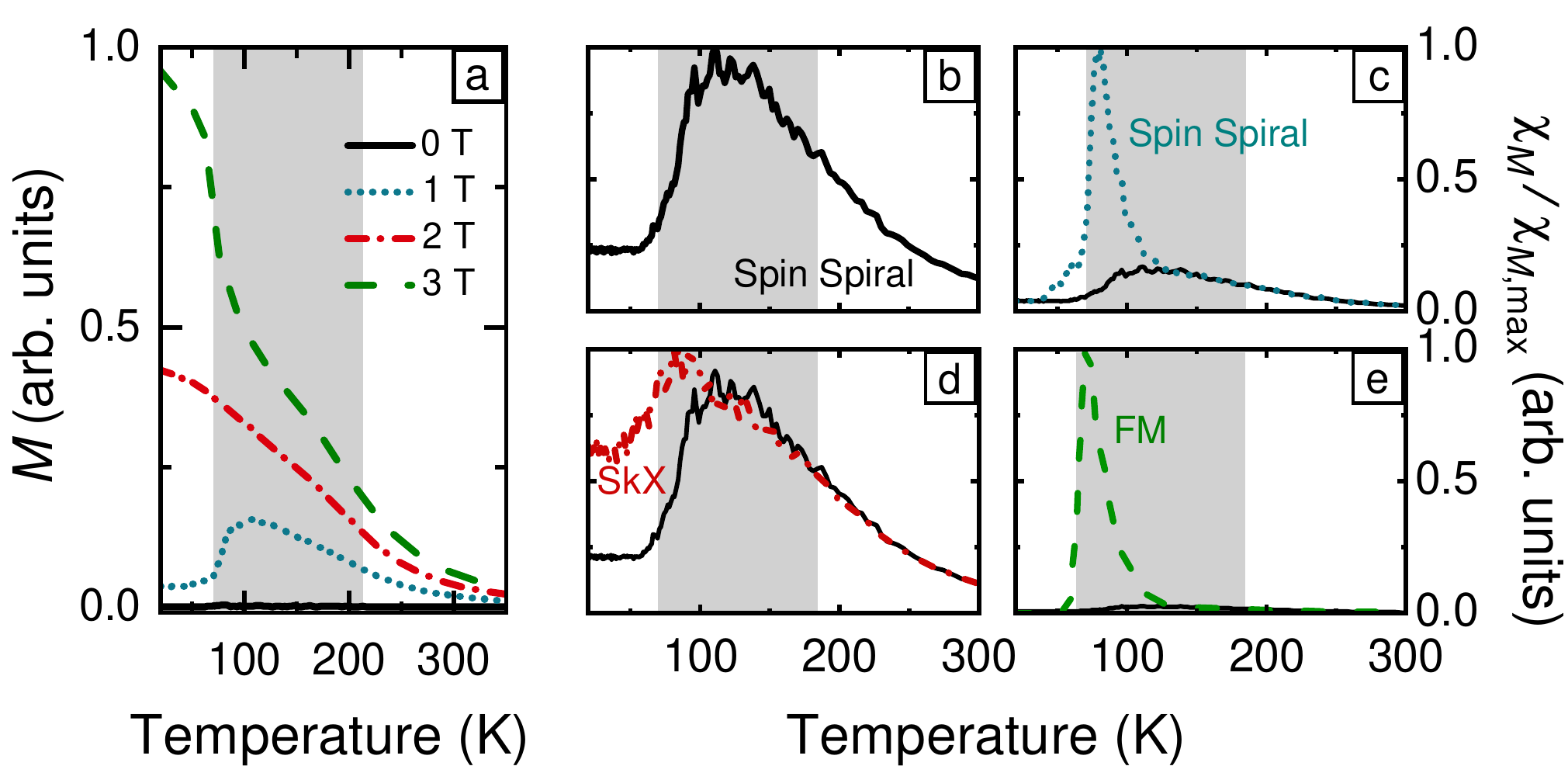}
\caption{{\bf Magnetization \textit{M} and magnetic susceptibility $\boldsymbol{\chi_M}$ at selected magnetic fields.} The fields are chosen in  the spin spiral ground state (full black line for $B$ = 0 T and dotted blue line for $B$ = 1 T), SkX ground state (chained red line for $B$ = 2 T) and FM ground state (dashed green line for $B$ = 3 T). 
(a) Magnetization vs. temperature.
(b)-(e) Magnetic susceptibility vs. temperature. The magnetic susceptibility without magnetic field ($B$ = 0 T) is replotted as full black line in each subplot. The gray shaded areas mark an intermediate region between the long-range ordered and PM phase as guide to the eye.
}
\label{fig:Magnetization}
\end{figure}

In a second step, we analyze the magnetization $M$ in figure~\ref{fig:Magnetization} (a) and magnetic susceptibility $\chi_M$ in figures~\ref{fig:Magnetization} (b) - (e) for selected magnetic fields with different ground states according to the stability diagram in figure~\ref{fig:stabilitydiagram_Pd-Fe-Ir111}. For comparison, the magnetic susceptibility without magnetic ($B$ = 0 T) field is replotted in figures~\ref{fig:Magnetization} (b) - (e).

The behavior of the magnetization depends on the ground state (see figure~\ref{fig:Magnetization} (a)).
Without a magnetic field ($B$ = 0 T), the magnetization $M$ is zero due to the spin spiral structure. For that case, the magnetic susceptibility shows a broad peak similar to the one reported in R\'{o}zsa {\it et al.} for Pd/Fe/Ir(111) \cite{rozsa2016} and Buhrandt and Fritz for MnSi \cite{PhysRevB.88.195137}.

At low magnetic field in the spin spiral ground state regime ($B$ = 1 T), the magnetization (dotted blue line in figure~\ref{fig:Magnetization} (a)) is non-zero even at $T$ = 0 K, since the spin spiral rotates inhomogeneously due to the applied magnetic field. 
At $T$ $\approx$ 81 K, the magnetization increases steeply and then decreases with temperature. 
In the magnetic susceptibility (figure~\ref{fig:Magnetization} (c)), we find an additional sharp peak at the position of the increase of magnetization.

At magnetic fields with a SkX ground state, the magnetization is non-zero at low temperature and decreases smoothly with temperature (chained red line in figure~\ref{fig:Magnetization} (a)).
For this case, we find a broad peak in the magnetic susceptibility (figure~\ref{fig:Magnetization} (c) and (e)).

In the case of a FM ground state, at low temperature the system is fully magnetized (dashed green line in figure~\ref{fig:Magnetization} (a)). At about $T$ $\approx$ 71 K the magnetization decreases steeply.
In the magnetic susceptibility, we also find an additional peak at the low temperature side of the broad peak as in the spin spiral case with magnetic field (see inset figure~\ref{fig:Magnetization} (d)). 

\subsection{Topological charge}

The above analysis of the thermodynamical quantities indicate an additional intermediate region within the ordered phase. We shade this region in gray in figures~\ref{fig:Magnetization} (a) - (e) for guidance.

\begin{figure}[tp]
\includegraphics[width=0.9\linewidth]{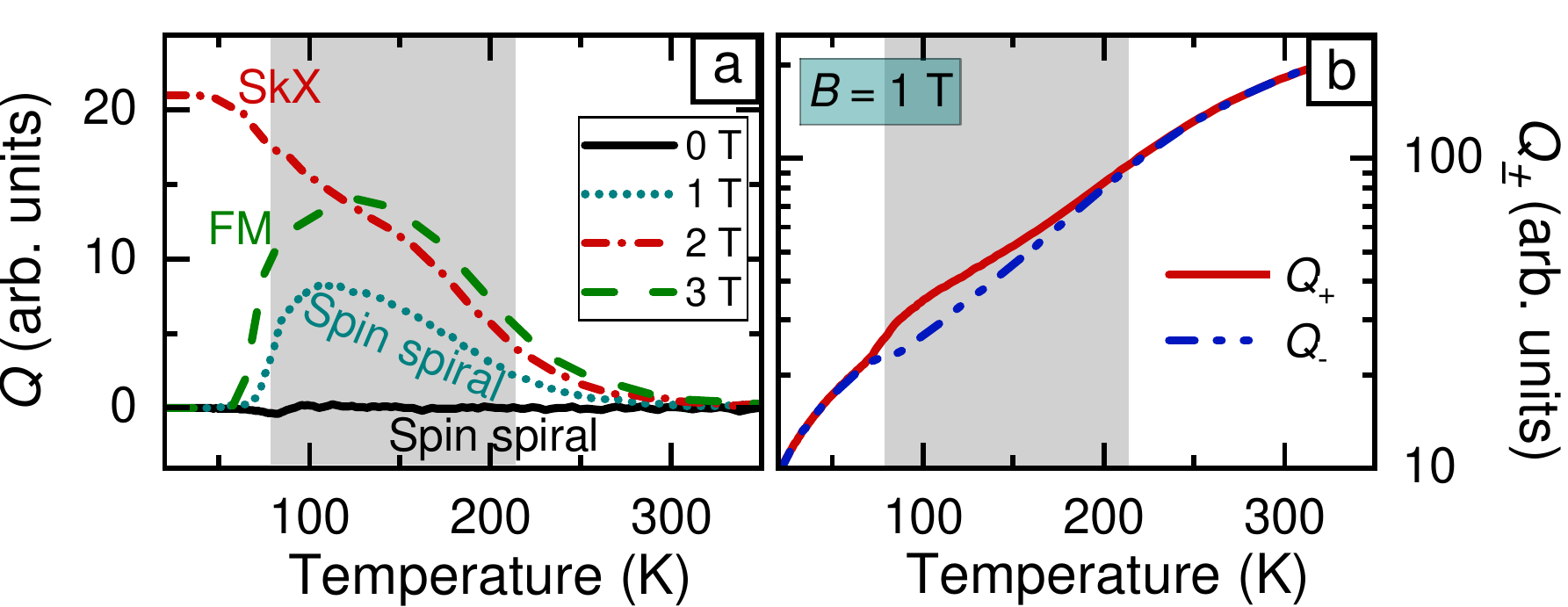}
\centering
\caption{{\bf Topological charge and corresponding SkD and ASkD contributions vs. temperature.} 
(a) Topological charge $Q$ vs. temperature for different magnetic fields. The fields are chosen in the spin spiral ground state (full black line for $B$ = 0 T and dotted blue line for $B$ = 1 T), SkX ground state (chained red line for $B$ = 2 T) and FM ground state (dashed green line for $B$ = 3 T). 
(b) SkD (red, full) and ASkD (blue, dashdotted) contributions $Q^+$ and $Q^-$, respectively vs. temperature at $B=1$ T.
The shaded gray area indicate the intermediate region for guidance.
}
\label{fig:TopologicalCharge}
\end{figure}

At first, we analyze the topological charge $Q$ in figure~\ref{fig:TopologicalCharge} (a).
Without magnetic field, $Q$ remains zero. This means, the DMI itself does not create the net topological charge and it needs the magnetic field to break the symmetry in agreement with Hou \textit{et al}. \cite{Hou2017a}.
In the case of a SkX ground state, $Q$ is maximum at low temperature and decreases smoothly with temperature.
When the magnetic fields stabilize the spin spiral or the FM ground states, $Q$ shows a similar trend when the temperature increases. In both cases, the topological charge increases steeply at a certain temperature and then decreases smoothly again. Therefore, the net topological charge cannot distinguish between a spin spiral and a FM ground state. Furthermore, it is not a good order parameter since the system may contain finite coexisting SkD and ASkD contributions $Q_+$ and $Q_-$, respectively \cite{Dupe2016a}. We therefore analyze in a second step the temperature dependence of both SkD and ASkD contributions.
%
%

The SkD and ASkD contributions $Q_\pm$ at $B$ = 1 T (spin spiral ground state) are shown in figure~\ref{fig:TopologicalCharge} (b).
At low temperature, the increase of the SkD (full red line) and ASkD (dashdotted blue line) contributions do not differ which means that the topological charge $Q$ remains constant. At high temperature, the SkD and ASkD are equal which is characteristic of the PM phase.
However, in a certain temperature range (shaded gray area), the SkD is larger than the ASkD.
Due to this difference, the topological charge is non-zero in this temperature range.

At magnetic fields with a FM ground state (not shown), the curves for the SkD and ASkD look similar to the ones of the spin spiral ground state.
In the case of a SkX ground state (not shown), both densities already differ at $T$ = 0 K, since the SkD is non-zero due to the SkX. 


\begin{figure}[thp]
\centering
\includegraphics[width=0.9\linewidth]{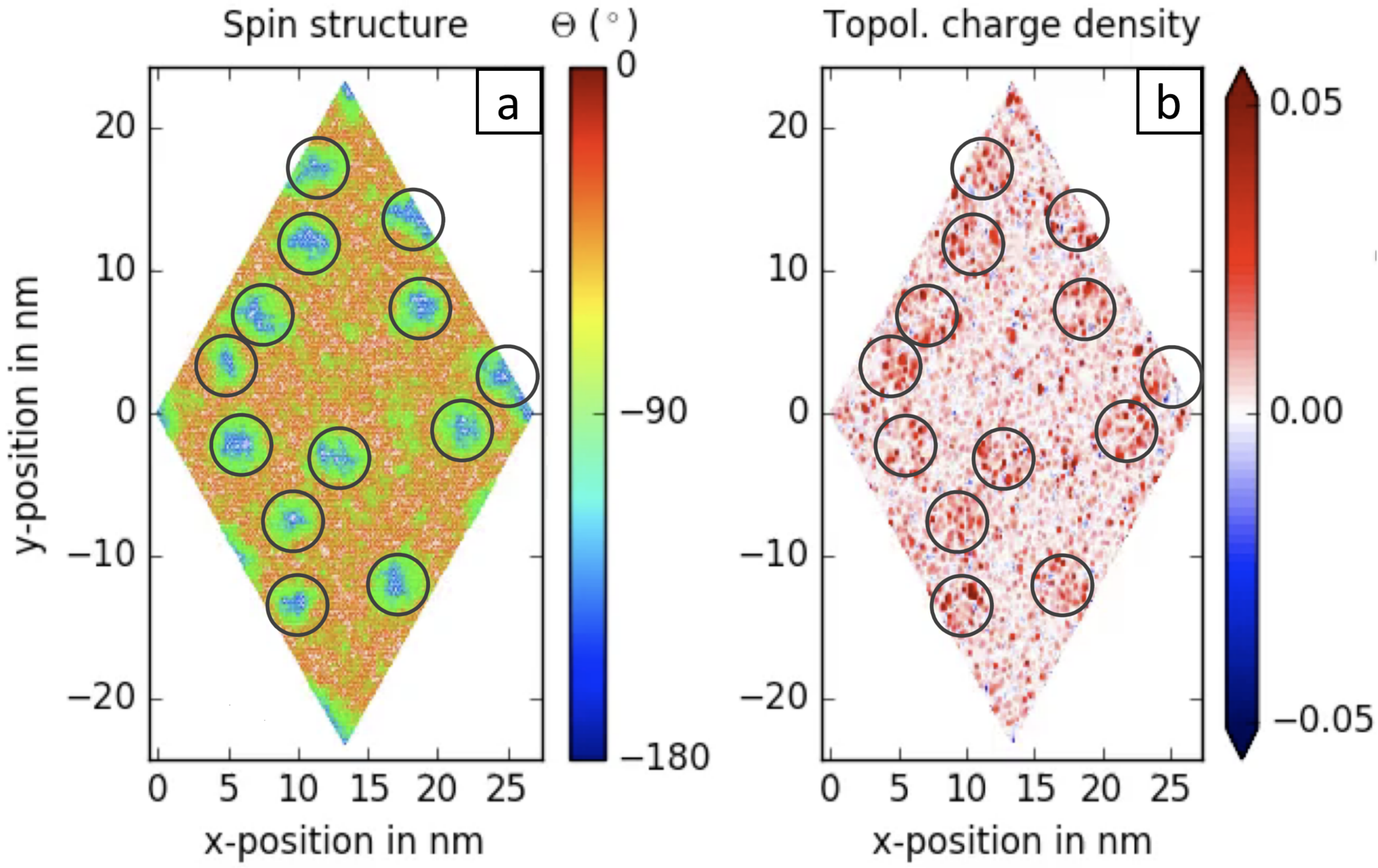}
\caption{{\bf Magnetic structure and distribution of the topological charge at 
\textit{B} = 3 T and \textit{T} = 72 K.
} The net topological charge is $Q$ = 13.
(a) Magnetic structure. The color code of the magnetization goes from red (out of the plane) over green (in plane) to blue (into the plane). The black circles indicate skyrmion-like magnetic structures.
(b) Spacial distribution of the SkD (red) and ASkD (blue). The black circles indicate accumulation of positive topological charge at the same position as the skyrmion-like magnetic structures in (a).
}
\label{fig:StructureAndCharge}
\end{figure}

To visualize the intermediate region, we show in figure~\ref{fig:StructureAndCharge} (a) snapshot of the magnetic structure and the distribution of the topological charge at the steep increase of the net topological charge at $T$ = 72 K for $B$ = 3 T. The magnetic structure shows 13 skyrmion-like patches (see figure~\ref{fig:StructureAndCharge} (a)) in a magnetized background. All patches are more or less spherically symmetric and have similar sizes.
The calculated net topological charge for this structure is $Q$ = 13. The spacial distribution of the corresponding SkD and ASkD can be seen in figure~\ref{fig:StructureAndCharge} (b). At the same position as the skyrmion-like patches in the magnetic structure, an accumulation of highter SkD is visible. We also find small ASkD contributions in the region of the patches, and both SkD and ASkD in the background, which would not be the case for condensed skyrmions in a FM background at very low temperature.
In supplementary videos, we show the evolution of these magnetic structures and spacial distributions of the SkD and ASkD as a function of Monte Carlo steps for selected magnetic fields and temperatures to show that their modification changes $Q$.


\begin{figure}[thp]
\centering
\includegraphics[width=0.8\linewidth]{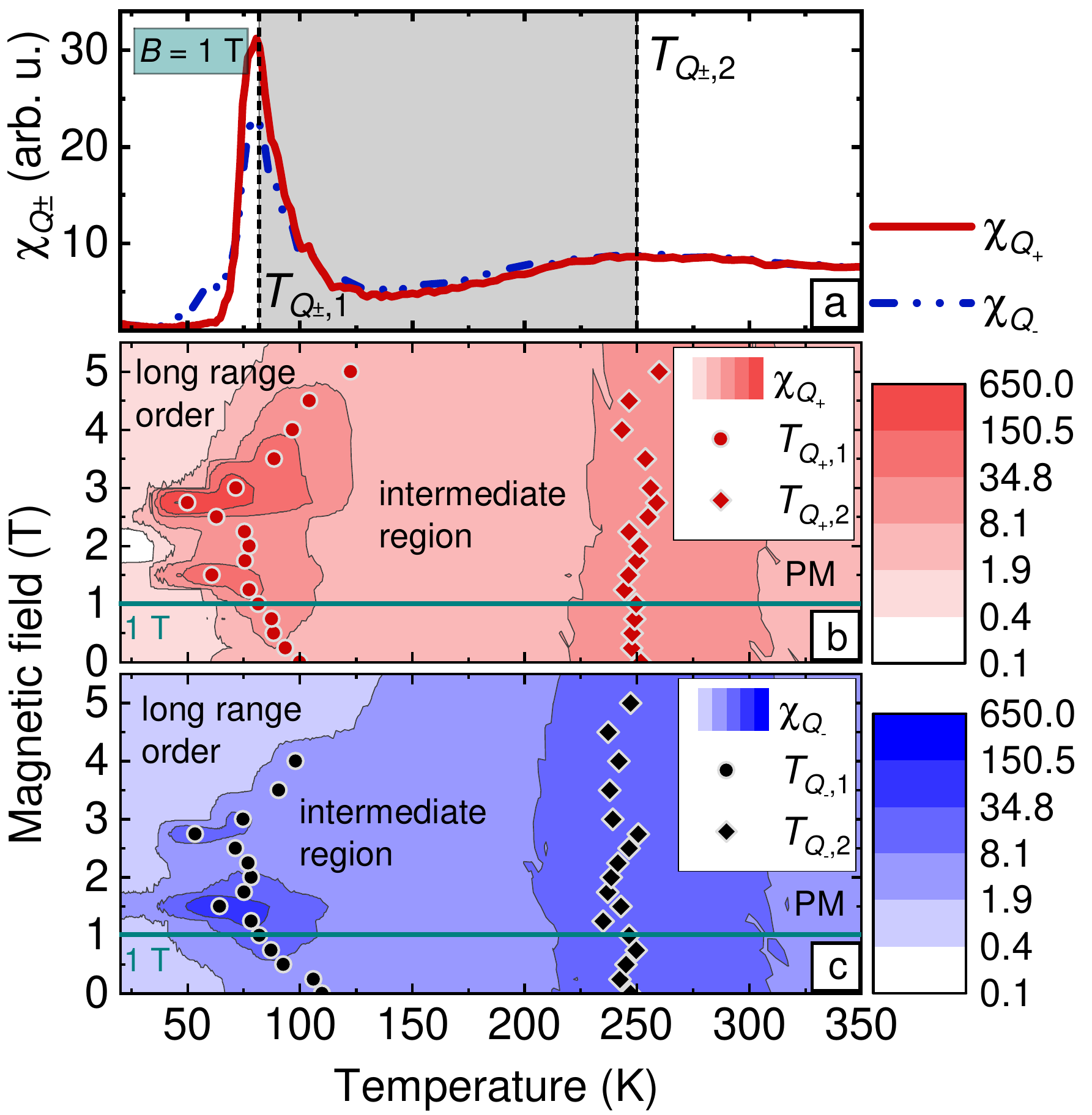}
\caption{{\bf Fluctuations of the topological charge densities SkD and ASkD.}
(a) Example of the topological susceptibilites $\chi_{Q_{\pm}}$ vs. temperature at $B$ = 1 T. By fitting the peaks with a lorentzian function, the critical temperatures $T_{Q_\pm,1}$ and $T_{Q_\pm,2}$ can be obtained. The gray shaded area in between these critical temperatures indicate the intermediate region.
(b) and (c) Heat map of the positive and negative topological susceptibility, respectively. The darker the areas the stronger are the fluctuations of the topological charge. The points and squares are the critical temperatures $T_{Q_\pm,1}$ and $T_{Q_\pm,2}$, which indicate the transition in the intermediate and PM region, respectively.
}
\label{fig:FluctuationsQ}
\end{figure}

To analyze the fluctuations of the topological charge densities SkD and ASkD, we calculate the topological susceptibilities $\chi_{Q_{\pm}}$ which are shown in figure~\ref{fig:FluctuationsQ}. 
We identify the transition temperatures $\chi_{Q_\pm,1}$ and $\chi_{Q_\pm,2}$, as the peak positions of the topological susceptibilities at fixed magnetic fields. 
As an example, we plot the topological susceptibilities $\chi_{Q_\pm}$ for $B$ = 1 T in figure~\ref{fig:FluctuationsQ} (a).
We find a broad peak with its maximum at about $T_{Q_\pm,2}$ $\approx$ 250 K. This peak indicates the transition from the intermediate region to the PM phase, as in the simple Heisenberg model in Ref.~\cite{PhysRevB.39.7212}.
However, at $T_{Q_\pm,1}$ $\approx$ 81 K an additional sharper peak appears, which indicates the transition from the ordered phase to the intermediate region.
In figure~\ref{fig:FluctuationsQ} (b) and (c) we plot a heat map of these topological susceptibilities at different temperatures and $B$ fields.
At high temperature, both topological susceptibilities have a maximum at all magnetic fields, indicating the transition to the PM phase. At low temperature, both susceptibilities show an asymmetric behavior. 
At magnetic fields with a spin spiral ground state, SkD and ASkD susceptibility are increasing symmetrical with magnetic field. This does not mean, that the same amount of SkD and ASkD is created, but that the fluctuations have the same correlation length and that both skyrmions and antiskyrmions may appear.

\subsection{Effect of magnetic exchange frustration on topological charge $Q$}

\begin{figure}[thp]
\centering
\includegraphics[width=0.9\linewidth]{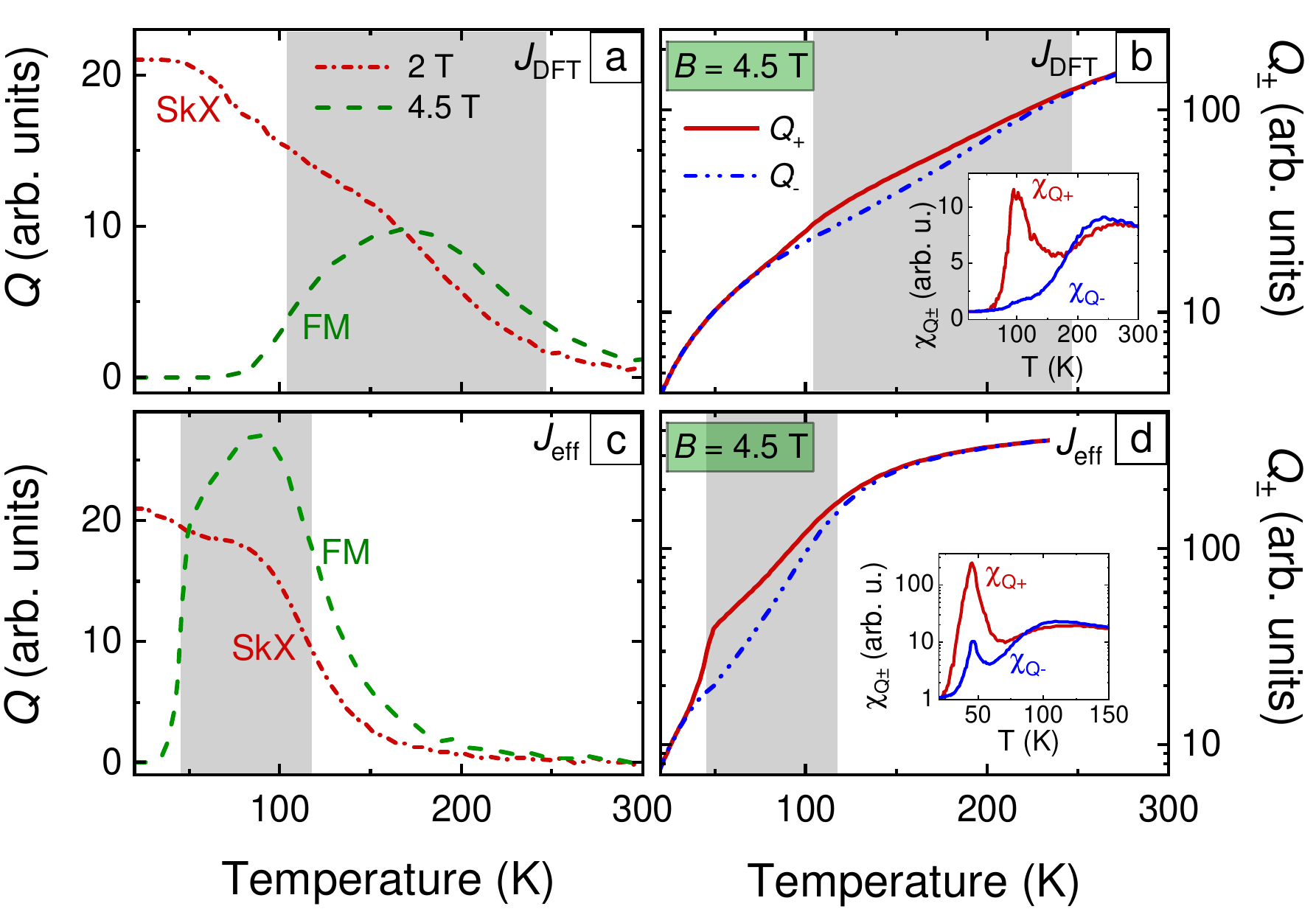}
\caption{{\textbf{Effect of frustration of exchange on the topological charge.}}
(a) Topological charge $Q$ vs. temperature for $J_\mathrm{DFT}$ at $B$ = 2~T (red chained line, SkX ground state) and $B$ = 4.5~T (green dashed line, FM ground state). The maximum of topological charge is in the region of the SkX
(b) SkD $Q_+$ (red, full) and ASkD $Q_-$ (blue, dashdotted) vs. temperature for $J_\mathrm{DFT}$ at $B$ = 4.5~T. The inset shows the corresponding topological susceptibilities $\chi_{Q_\pm}$.
(c) Topological charge $Q$ vs. temperature for $J_\mathrm{eff}$ at $B$ = 2~T (red chained line, SkX ground state) and $B$ = 4.5~T (green dashed line, FM ground state). The maximum of topological charge is in the intermediate region at a $B$ field with FM ground state. 
(d) SkD $Q_+$ (red, full) and ASkD $Q_-$ (blue, dashdotted) vs. temperature for $J_\mathrm{eff}$ at $B$ = 4.5~T. The inset shows the corresponding topological susceptibilities $\chi_{Q_\pm}$.
The gray shaded areas indicate the intermediate region.}
\label{fig:effectiveExchangeParameter}
\end{figure}

We now analyze the effect of the frustration of exchange interaction on the topological charge and the topological susceptibility. We therefore performed PTMC simulations at different magnetic fields with only an effective exchange coefficient $J_\mathrm{eff}$ with 192 temperature steps. These parameters were derived in Ref.~\cite{vonMalottki2017} and give a similar stability diagram as in figure~\ref{fig:stabilitydiagram_Pd-Fe-Ir111}. They are also very similar to those obtained based on fitting experimental data to the micromagnetic model \cite{Romming2015}.

In figures~\ref{fig:effectiveExchangeParameter} (a) and (c) is plotted the topological charge $Q$ for $J_\mathrm{DFT}$ and $J_\mathrm{eff}$, respectively, for $B$ = 2 T (magnetic field with SkX ground state) and $B$ = 4.5~T (magnetic field with FM ground state). It shows, that in the case of a frustration of exchange ($J_\mathrm{DFT}$), the maximum is in the SkX, whereas for the case of only an effective exchange parameter ($J_\mathrm{DFT}$), the maximum of topological charge is in the intermediate region at a magnetic field with FM ground state. 

The figures~\ref{fig:effectiveExchangeParameter} (b) and (d) show the positive and negative topological charge contributions $Q_\pm$ for $J_\mathrm{DFT}$ and $J_\mathrm{eff}$, respectively, for $B$ = 4.5~T (magnetic field with FM ground state). In both cases, the topological charges do not differ at low and high temperature, but in the intermediate region the positive charge exceeds the negative one. The inset plots show the corresponding topological susceptibilities $\chi_{Q_\pm}$. In the case of frustrated exchange ($J_\mathrm{DFT}$), we identify a broad peak indicating a transition to the PM phase for both charges at about $T_{Q_\pm,2}$ = 260~K as in Ref.~\cite{PhysRevB.39.7212} and for the positive topological charge also an additional peak indicating the transition to the intermediate region at about $T_{Q_+,1}$ = 100~K. However, there is no peak indicating a transition to the intermediate region for the negative charge, of which the slope is smooth as a function of temperature.
In the case of an effective exchange parameter ($J_\mathrm{eff}$), both charges show a broad peak at about $T_{Q_\pm,2}$ = 115~K, and here also both charges show a peak indicating the transition to the intermediate region at about $T_{Q_\pm,1}$ = 45~K, whereby the peak of the positive charge is much larger than the negative one.
Both peaks indicate the transition to the intermediate region ($T_{Q_\pm,1}$).

The above results could indicate, that the frustration of exchange yield different chiral excitations.
In the case of a frustration of exchange, the magnons are driven by exchange, and in the case of an effective exchange parameter driven by DMI, which would favor shorter wave length excitations in the intermediate region and 
unfavor the creation of negative topological charge in agreement with the calculated energy barriers for antiskyrmion collapse of Ref.~\cite{vonMalottki2017} 
When using $J_\mathrm{DFT}$, the magnetic exchange interactions favor magnons of 6-7~nm periodicity and the SkX is therefore the maximum of $Q$.

\subsection{$B$-$T$ phase diagram}

\begin{figure}[thp]
\centering
\includegraphics[width=0.9\linewidth]{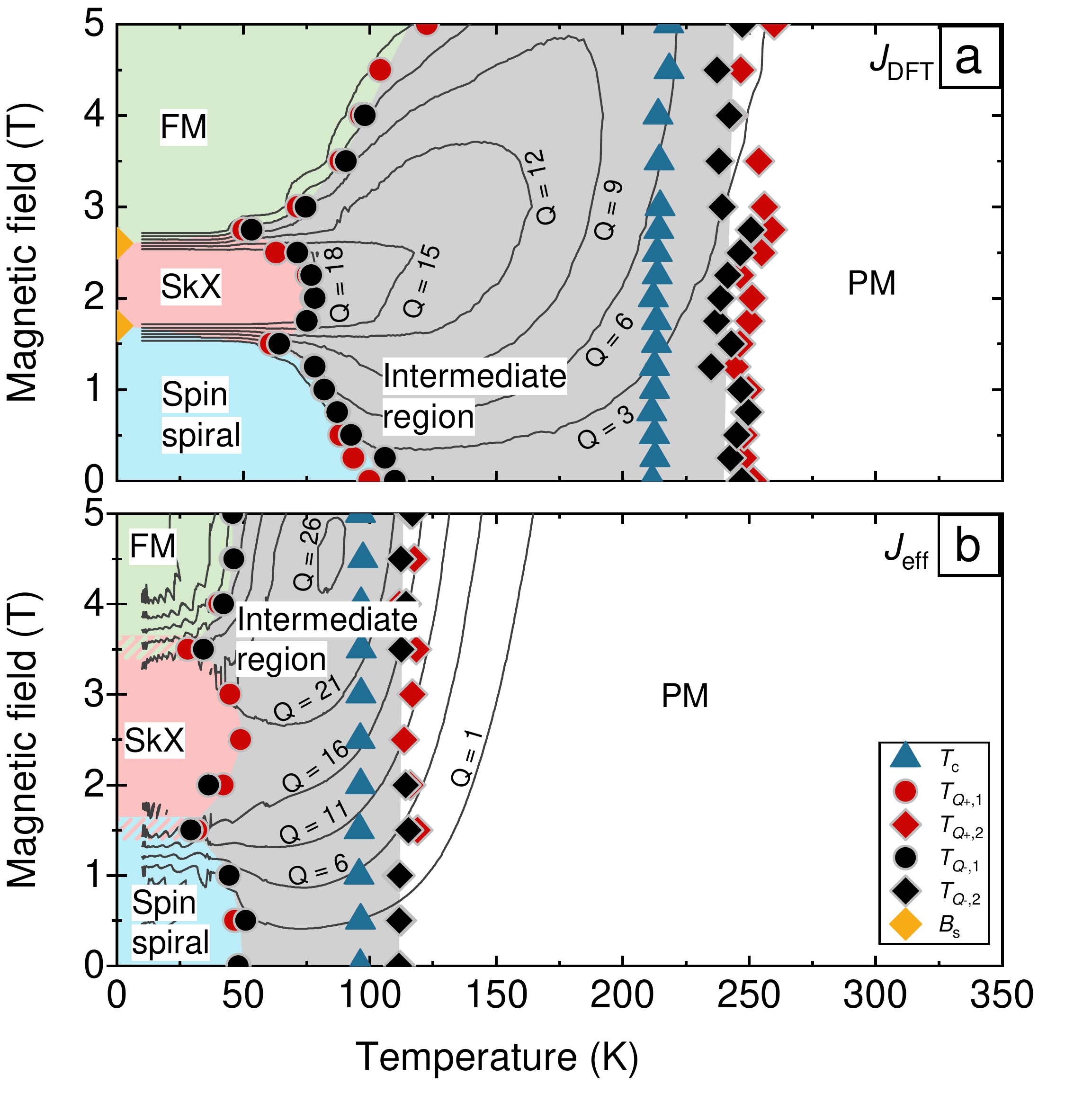}
\caption{{\textbf{\textit{B}-\textit{T} phase diagram of Pd/Fe/Ir(111)}}
In (a) the full set of exchange constants of DFT which include the frustration and (b) only an effective exchange parameter is taken into account.
Five different phases are identified (shaded for guidance): spin spiral region (blue), SkX region (red), FM phase (green), intermediate region (gray) and PM phase. The contour lines show the number of net topological charge.
}
\label{fig:Phasediagram_Color}
\end{figure}

We summarize the information of the transition temperatures $T_\mathrm{c}$, $T_{Q_\pm,1}$ and $T_{Q_\pm,2}$ in a $B$-$T$ phase diagram which is shown in figure~\ref{fig:Phasediagram_Color} (a) for the case of frustration of exchange ($J_\mathrm{DFT}$), where we plot the mean topological charge $Q$ as contour lines.
In total we identified five different phases.
At low temperature three ordered phases occur with increasing magnetic field: spin spiral, SkX and FM phase.
At higher temperatures ($T_{Q_\pm,1}$) the system shows a transition into the intermediate region, with a non-zero topological charge $Q$.
In contrast to a previously reported phase diagram \cite{rozsa2016}, we find critical temperatures which depend on the magnetic field.
Here, we find a decrease of the critical temperature with increasing magnetic field in the region with a spin spiral ground state.
When the FM state is lowest in energy, the critical temperatures increase with increasing magnetic field. 
With further increase of the temperature, the topological charge vanishes and the system becomes PM.
The critical temperature of this transition to the PM state can be defined as the position of the peak of the heat capacity ($T_{\textup{c}}$) or 
the broad peak of the topological susceptibility ($T_{Q_\pm,2}$).

As a last step, we calculate the $B$-$T$ phase diagram when only an effective exchange parameter is taken into account ($J_\mathrm{eff}$) which is shown in figure ~\ref{fig:Phasediagram_Color} (b).
The magnetic structures of the different ground states do not differ from the ones in figure~\ref{fig:stabilitydiagram_Pd-Fe-Ir111}, the different regimes overlap at low temperature and the $B$-$T$ phase diagram shows again five phases (spin spiral, SkX, FM phase, intermediate region, PM phase). However the critical temperatures are reduced by a factor of roughly 2, compared to the case, when frustration of exchange is included (see figure ~\ref{fig:Phasediagram_Color} (a)). Furthermore, the maximum of topological charge is not in the SkX phase as in the case for a full set of exchange coefficients, but in the intermediate region with a magnetic field which stabilizes a FM ground state.

\section{Conclusion}

We could show, that at low temperatures, the critical temperature from the long range ordered phase to an intermediate region decreases in the case of a spin spiral ground state and increases in case of a FM ground state with increasing magnetic field.
We showed that the pairwise creation of SkD and ASkD are good quantities to study the critical temperatures and chiral excitations.
In the ordered phases, both the SkD and ASkD distribution are created at the same rate. In the intermediate region, the SkD increases faster than the ASkD which results in a net topological charge.
At fixed magnetic field, in a certain temperature range, SkD and ASkD is created asymmetrically which yields a net topological charge even at magnetic fields without a SkX ground state.
We showed, that the critical temperatures are increased by roughly a factor of 2, when frustration of exchange is taken into account. In addition, we find that the maximum of topological charge is in the SkX lattice phase for the full set of exchange coefficients and in contrast to that in the intermediate region with FM ground state, when only an effective exchange parameter is taken into account. 
We interpret these differences as the occurrence of different metastable states, depending on the stabilization mechanism.
Our simulations provide quantities that could be measured that would distinguish between the two models.

\ack 
The authors would like to gratefully acknowledge K Binder for insightful discussions and careful reading of the manuscript.
The authors gratefully acknowledge insightful discussions with M Garst and S Buhrandt.

This research was supported by DFG project DU 1489/2-1, the Alexander von Humboldt Foundation, the Graduate School Materials Science in Mainz, the Transregional Collaborative Research Center (SFB/TRR) 173 SPIN+X. J Sinova thanks the Grant Agency of the Czech Republic grant no. 14-37427G and the ERC Synergy Grant SC2 (No. 610115). M B\"ottcher and S Heinze thank the European Unions Horizon 2020 research and innovation program under grant agreement No. 665095 (FET - Open project MAGicSky).

B Dup\'e, S Heinze and M B\"ottcher gratefully acknowledge computing time at the HLRN and Mogon supercomputers.

\section*{References}
\bibliography{MaLit}

\end{document}